\documentclass[aip,amsmath,amssymb,reprint]{revtex4-1}

\usepackage{hyperref}
\usepackage{graphicx}
\usepackage{dcolumn}
\usepackage{bm}

\usepackage[utf8]{inputenc}
\usepackage[T1]{fontenc}
\usepackage{mathptmx}

\begin{document}

\preprint{AIP/123-QED}

\title{Elastic properties of few unit cell thick superconducting crystals of Bi$_2$Sr$_2$CaCu$_2$O$_{8+\delta}$}

\author{Sudhir~Kumar~Sahu}
\affiliation{Department of Physics, Indian Institute of Science, Bangalore-560012 (India)}
\author{Digambar~Jangade}
\affiliation{Department of condensed matter physics and material sciences, Tata Institute of Fundamental Research, Mumbai - 400005 (India)}
\author{Arumugam~Thamizhavel}
\affiliation{Department of condensed matter physics and material sciences, Tata Institute of Fundamental Research, Mumbai - 400005 (India)}
\author{Mandar~M.~Deshmukh}
\affiliation{Department of condensed matter physics and material sciences, Tata Institute of Fundamental Research, Mumbai - 400005 (India)} 
\author{Vibhor~Singh}
\email{v.singh@iisc.ac.in}
\affiliation{Department of Physics, Indian Institute of Science, Bangalore-560012 (India)}

\date{\today}

\begin{abstract}
We present systematic measurements of the mechanical properties of few unit cell (UC) thick
exfoliated crystals of a high-T$_c$ cuprate superconductor Bi$_2$Sr$_2$CaCu$_2$O$_{8+\delta}$. 
We determine the elastic properties of these crystals by deformation using an atomic 
force microscope (AFM) at room temperature. 
With the spatial 
measurements of local compliance and their detailed modeling, we independently determine the Young's 
modulus of rigidity and the pre-stress. The Young's modulus of rigidity is found to be in the 
range of 22~GPa to 30~GPa for flakes with thickness from $\sim$ 5~UC to 18~UC. The pre-stress 
spreads over the range of 5~MPa - 46~MPa, indicating a run-to-run 
variation during the exfoliation process. The determination of Young's modulus of rigidity for 
thin flakes is further verified from recently reported buckling technique. 
\end{abstract}

\maketitle

There has been a keen interest towards the use of two-dimensional (2D) 
thin materials for device applications.  While the electrical properties 
of these materials cover a wide spectrum ranging from insulating, 
semiconducting, metallic to superconducting behavior, their mechanical 
properties such as modulus of rigidity, fracture-strain, and thermal 
expansion are equally intriguing \cite{akinwande_two-dimensional_2014,castellanos-gomez_mechanics_2015}. 
This remarkable combination of characteristics make these materials accessible to novel applications 
such as flexible electronics and hybrid nano-electromechanical systems 
for sensing applications \cite{singh_optomechanical_2014,guttinger_force_2016,will_high_2017,sahu_nanoelectromechanical_2019}.

For nanoelectromechanical devices, materials with high electrical conductivity,
and low mass are often preferred,
as these properties tend to minimize the losses and improve the displacement transduction  \cite{singh_optomechanical_2014,guttinger_force_2016,will_high_2017,sahu_nanoelectromechanical_2019}. 
The mechanical properties of few unit cells (UC) thick exfoliated crystals could be significantly
different from its bulk counterpart, resulting in interesting effects such as nonlinear 
damping \cite{singh_negative_2016}, and Duffing phenomena \cite{davidovikj_nonlinear_2017}.
In addition, measurements of the elastic response could be a sensitive probe to the electronic or structural phase transitions in these materials \cite{binek_elastic_2017}. 
This has led to a 
considerable investigation into the nanomechanical properties of materials such as 
graphene, MoS$_2$, NbSe$_2$, etc.  \cite{poot_nanomechanical_2008,sengupta_electromechanical_2010,lee_measurement_2008,castellanos-gomez_elastic_2012,falin_mechanical_2017,liu_elastic_2014}.

Recently, few UC thick crystals of high-transition temperature superconductor 
Bi$_2$Sr$_2$CaCu$_2$O$_{8+\delta}$ (BSCCO) have attracted attention due to 
their unique superconducting phase diagram, and applications towards cavity-optomechanical 
devices \cite{huang_reliable_2015,sterpetti_comprehensive_2017,liao_superconductorinsulator_2018,sahu_nanoelectromechanical_2019}.
While the elastic coefficients of the bulk crystals of BSCCO have been 
observed with large variations, there is no investigation into the
elastic properties of few UC thick nanoscale samples \cite{sihan_elastic_1989,nes_anomalies_1991}.
Here we report the measurement of Young's modulus of rigidity $(E)$ and pre-stress $(\sigma)$ on few UC thick superconducting 
crystals of BSCCO. These properties are helpful in engineering the resonant frequency of 
mechanical resonators for composite devices. In addition, determination of the Young's modulus of rigidity by two different methods, on the crystals grown in the same run, brings clarity to the previously reported results from bulk-crystals \cite{sihan_elastic_1989,nes_anomalies_1991}.

We have primarily used elastic deformation by an AFM tip to measure the Young's 
modulus of rigidity and the pre-stress in exfoliated flakes of BSCCO. 
In total, we have studied 7 mesoscopic samples having thickness in the range of 16~nm to 
55~nm, corresponding to $\sim$5~UC to 18~UC thick crystals. We have performed measurements of the local compliance of the suspended 
flakes of BSCCO. Detailed finite-element modeling is then carried out 
to extract the Young's modulus and the pre-stress from compliance measurements. 
In addition, we employ a recently reported buckling technique to independently measure the Young's modulus
for thin flakes \cite{stafford_buckling-based_2004}. We find that the results from the buckling technique are consistent with the AFM technique.

\begin{figure}
	\begin{center}
		\includegraphics[width = 60 mm]{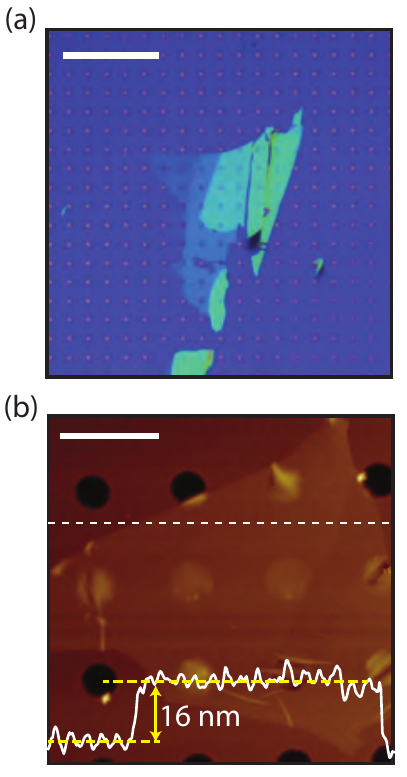}
		\caption{(a) An optical microscope image of BSCCO flake on 
			top of a patterned Si substrate coated with 285~nm of SiO$_2$. Difference 
			in the contrast for suspended and collapsed drums can be seen. 
			The scale-bar corresponds to 50~$\mu$m. (b) AFM image showing the 
			topography of the flake. A height profile taken at the position marked by a dashed line is overlaid on the AFM image, showing a thickness of $\sim$16~nm. The scale bar corresponds to 
			8~$\mu$m. }\label{Fig1}
	\end{center}
\end{figure}

High-quality single-crystals of BSCCO were prepared by annealing melt-quenched 
shards in oxygen atmosphere. The BSCCO shards were prepared by heating the BSCCO powder (Sigma Aldrich - 365106) in a high-quality recrystallized alumina crucible. After the annealing step, these crystals are stored 
in liquid nitrogen and are only taken out at the time of mechanical exfoliation \cite{jindal_growth_2017}.
To prepare the samples for AFM measurements, we first use photolithography 
to pattern circular trenches in photoresist (S1813) 
on a 285~nm thick SiO$_2$ coated silicon substrate. It is followed by a 
step of reactive-ion etching of SiO$_2$ using a low-pressure fluorine plasma. 
This results in circular trenches having diameter from 1.8 to 6~$\mu$m on the substrate.
Thin flakes of BSCCO are exfoliated by a scotch tape and transferred
on top of the patterned substrate using a polydimethylsiloxane (PDMS) based dry exfoliation technique 
\cite{castellanos-gomez_deterministic_2014}.

Fig.~\ref{Fig1}~(a) shows an optical image of a transferred BSCCO flake 
on the patterned substrate. Different optical contrast for suspended and collapsed 
micro-drums can be seen in the image. Fig.~\ref{Fig1}~(b) shows the topography of 
BSCCO flake alongside a height profile measured by AFM. Drums with an uneven 
topographic profile are not considered for measurements. The white dotted line 
indicates the location of the measured height profile. Transfer of flakes thinner 
than 16~nm ($\sim$ 5~UC) is a challenging task with our technique as they offer poor 
optical contrast, and their identification on PDMS remains difficult.


Elastic deformation by an AFM tip is a well-established method to characterize the elastic 
properties of nanoscale materials \cite{poot_nanomechanical_2008,lee_measurement_2008,norouzi_how_2006}.
An AFM cantilever with a known spring constant 
is used to apply a force on top of the suspended structure. This force deflects 
the flake depending on the elastic properties of the material and boundary conditions. In this study, 
we used an AFM tip with a spring constant of $k_{tip}$~=~5.6~N~m$^{-1}$, measured using thermo-mechanical
noise calibration \cite{sader_calibration_1999}. The spring constant of the tip relates the applied force $F$
to the tip deflection $\Delta z_{tip}$, given by $F = k_{tip}\Delta z_{tip}$. The elastic 
deformation of the flake $\delta$ then can be expressed in terms of the net displacement of 
the AFM piezo $\Delta z_p$ as: $\delta=\Delta z_p-\Delta z_{tip}$.

\begin{figure}
	\begin{center}
		\includegraphics[width = 65 mm]{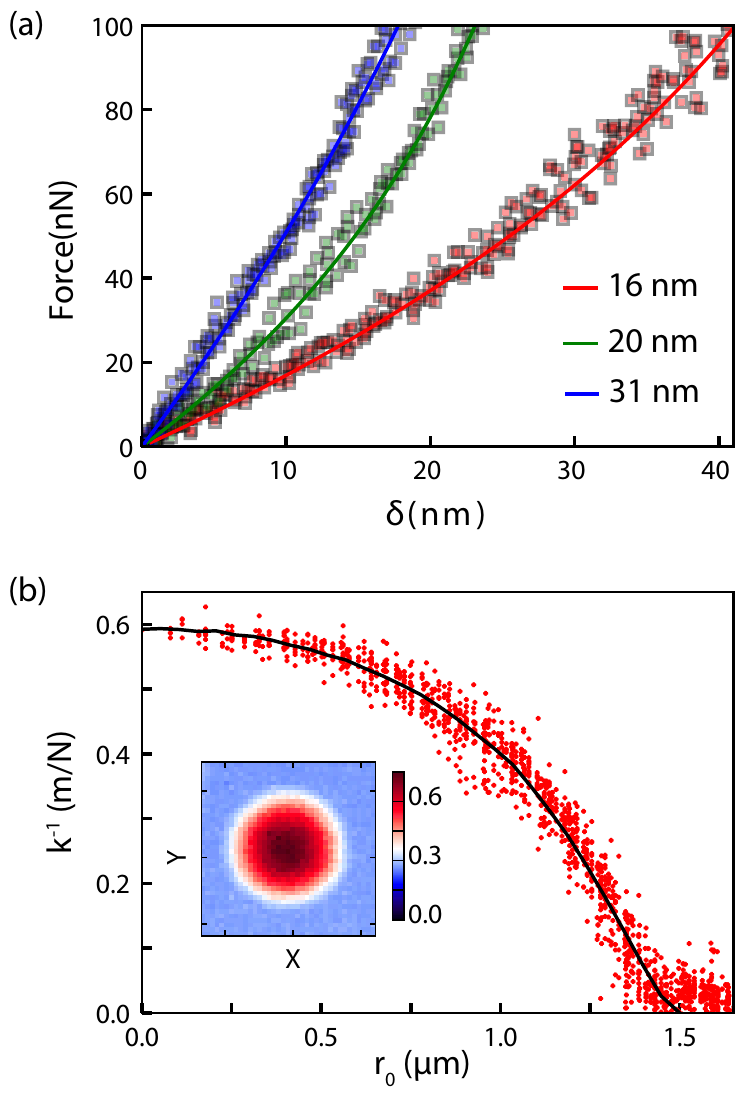}
		\caption{(a) Force-displacement curves and corresponding fits for 
			flakes with thicknesses of 16~nm (red), 20~nm (green), and 31~nm (blue). 
			(b) Volumetric measurement of the compliance for a flake with a 
			thickness of 16~nm. The solid line is fit to the data.  The inset shows the complete compliance map measured 
			from a flake of $\sim$ 3~$\mu$m diameter. Ticks on the inset figure are placed 1.5~$\mu$m 
			apart.
		}\label{Fig2}
	\end{center}
\end{figure}

Fig.~\ref{Fig2}~(a) shows traces of loading-curve for flakes of different thicknesses
while indenting at the center of the circularly-shaped suspended part. We do not observe 
any hysteresis between loading and unloading curves, suggesting no slippage at the flake boundary.
The loading curve is quite linear for the thick flake (31~nm). A nonlinear behavior can be easily 
observed for the thin flake (16~nm). Within continuum mechanics for an isotropic solid, the elastic deformation of the 
flake $\delta$ is related to the applied force $F$ as: 

\begin{equation}
	F=\left[{\frac{4 \pi E}{3(1-\nu ^2)}} \left({\frac{t^3}{a^2}}\right)\right]\delta+ 
	\left({\sigma\pi t}\right)\delta+\left({\frac {q^3 E t}{a^2}}\right)\delta^3,
	\label{eq2}
\end{equation}

where $q= 1/(1.049-0.15\nu-0.16\nu^2)$ is a dimensionless constant, $a$ and $t$ are the radius and thickness of the suspended BSCCO flake, respectively \cite{timoshenko_theory_1959}. By using a Poisson's ratio of $\nu$~=~0.2 for BSCCO \cite{sihan_elastic_1989}, and other geometrical quantities obtained from the AFM measurements, we fit the loading curves using Eq.~\ref{eq2}, shown by the solid lines in Fig.~\ref{Fig2}~(a).

In the membrane-limit, the contribution from bending rigidity (first term in Eq.~\ref{eq2}) can be neglected and the nonlinear relation between force and deformation can be used to extract the pre-stress and Young's modulus independently.
In the plate-limit however, due to the comparable contributions from bending 
rigidity and tensile stress to the total elastic energy, it is 
impossible to separate out $\sigma$ and $E$ from the deformation measurement at 
the center of the flake alone. Therefore, we resort to the spatial measurement of local compliance, defined as  
$k^{-1} (r_0, \theta_0) = \left(d\delta/dF\right)|_{r_0,\theta_0}$, over the
suspended part of the flake
\cite{radmacher_imaging_1994}. 
Spatial map of the measured local compliance 
over a grid of 64$\times$64 points for a suspended flake of $\sim$3~$\mu$m diameter 
is shown in the inset of Fig.~\ref{Fig2}~(b). As the compliance profile is radially 
symmetric, Fig.~\ref{Fig2}~(b) shows the variation of $k^{-1}$ with the distance 
from the center of the flake $r_0$.

\begin{figure*}
	\begin{center}
		\includegraphics[width = 160 mm]{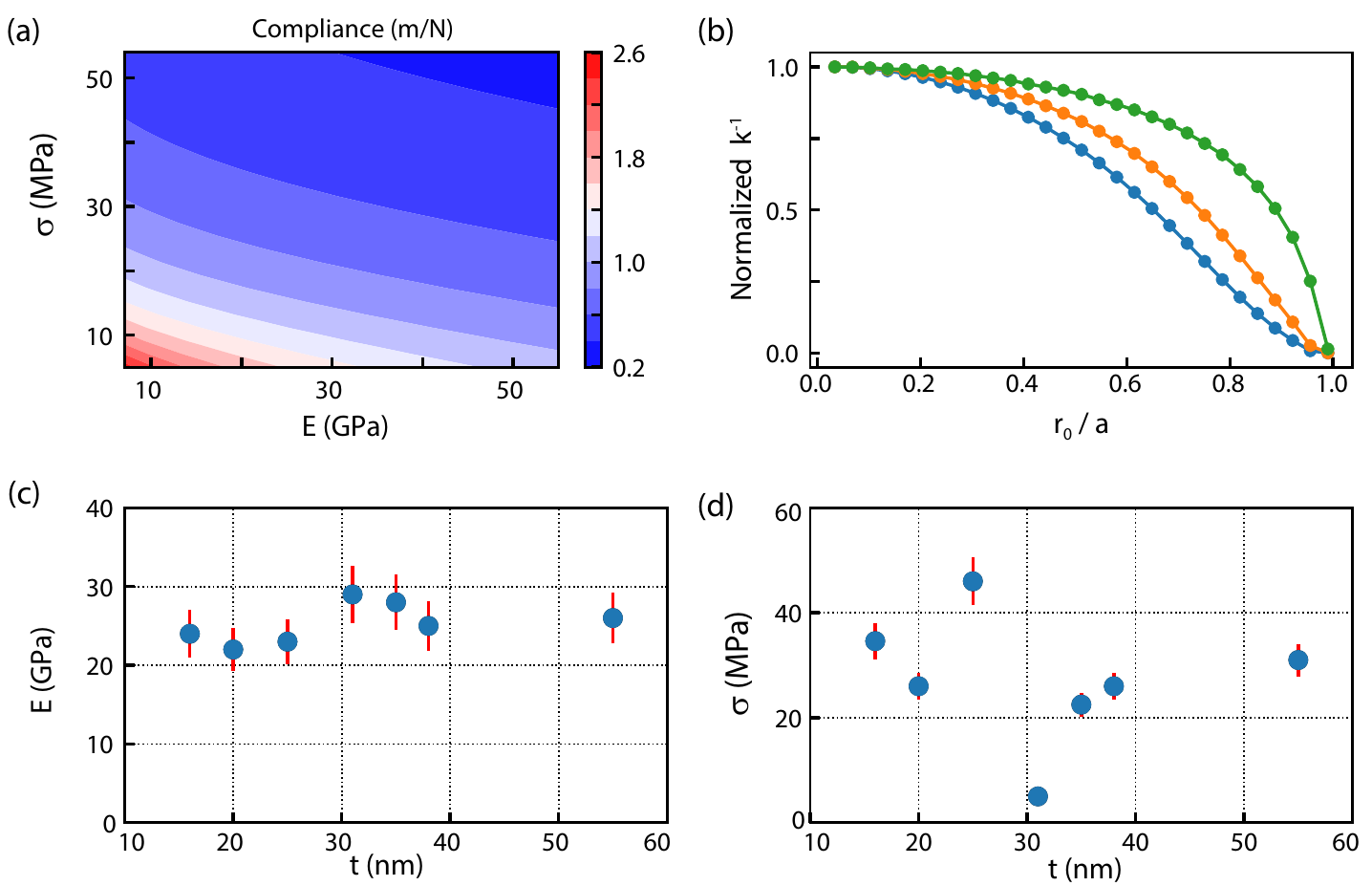}
		\caption{(a) Contour plot showing local compliance $k^{-1}$ at the center of the drum with a variation of the Young's modulus and pre-stress, simulated by finite element method. (b) The simulated radial profile of compliance for different values of $\lambda$ given by 1~(blue), 10~(orange) and 100~(green), respectively. Panels (c) and (d) show the extracted value of the Young's modulus and pre-stress of BSCCO flakes of different thicknesses.  The error bars are 
			calculated from the spread in the radial compliance data. }\label{Fig3}
	\end{center}
\end{figure*}

To extract the elastic properties of the BSCCO flakes from the spatial compliance 
maps, we perform finite element simulations using COMSOL (see Supplementary Material (SM) for details). For a linear elastic solid, 
the deformation under a point load can be well described by the Euler-Lagrange differential equation
\cite{landau_theory_1986}. However, when the contribution of bending rigidity and pre-stress 
to the elastic energy are comparable, it is difficult to find a closed-form 
solution of the elastic deformation $\delta(r,\theta)$ and hence compliance 
$k^{-1} = \partial \delta/\partial F$ \cite{norouzi_how_2006}. To model the system, 
we consider the deformation of a linear elastic material under a load 
applied by an AFM tip of 40~nm radius.
A rigid boundary condition 
is applied at the edge of the flake i.e. $\delta|_{r=a} = 0$. The sliding 
contact between the AFM tip and flake is captured by applying sliding contact 
boundary condition. 
Other material parameters such as the pre-stress
and the Young's modulus are supplied as inputs to the model. Thus, the calculation of 
deflection under a small load applied at point $r_0$ away from the center 
of the flake allows calculating the local compliance.

Fig.~\ref{Fig3}(a) shows a contour plot of simulated compliance at the center ($r_0=0$) of a 3~$\mu$m 
diameter flake for different values of $E$, and $\sigma$. It is obvious from the plot that different 
combinations of $(E, \sigma)$ can result in the same value of the compliance at the center of the flake. 
The radial shape of the compliance, however, depends on the ratio of pre-stress and bending rigidity 
$D~=~\frac{Et^3}{12(1-\nu^2)}$.
Fig.~\ref{Fig3}~(b) shows the plots of the simulated radial profile of normalized compliance 
$k^{-1}/k^{-1}(r_0=0)$ for three different values of $\lambda$, defined as $\lambda~=~\sqrt{\frac{\sigma t a^2}{D}}$.
To fit the simulated results with the experimental data, we choose the contour of $k^{-1}(r_0 = 0)$ 
that matches with the AFM data. Along this contour, radial compliance profiles are computed for various combinations 
of $(E, \sigma)$ to fit the experimentally obtained data. A result of
this procedure is shown in Fig.~\ref{Fig2}~(b) by a black continuous line.

A plot summarizing the Young's modulus and the pre-stress for 7 different exfoliated flakes  
of varying thickness is shown in Fig.~\ref{Fig3}~(c,~d). Detailed characterization of these flakes is 
provided in the SM. It is important to highlight 
that the pre-stress in these flakes results from the dry transfer process 
and is independent of material properties. Therefore, it spreads over significantly 
from 5~MPa to 46~MPa.  
However, the Young’s modulus of rigidity is found to be 
in the range of 22~GPa to 30~GPa.
Typically, elastic coefficients of ultra-thin samples, where the surface elastic 
energy is non-negligible to the bulk elastic energy, show a 
thickness-dependence  \cite{miller_size-dependent_2000,lee_measurement_2008,falin_mechanical_2017}. 
We do not observe any prominent thickness dependence in the Young's modulus of rigidity as the samples studied here are at least 16~nm thick.

\begin{figure}
	\begin{center}
		\includegraphics[width = 65 mm]{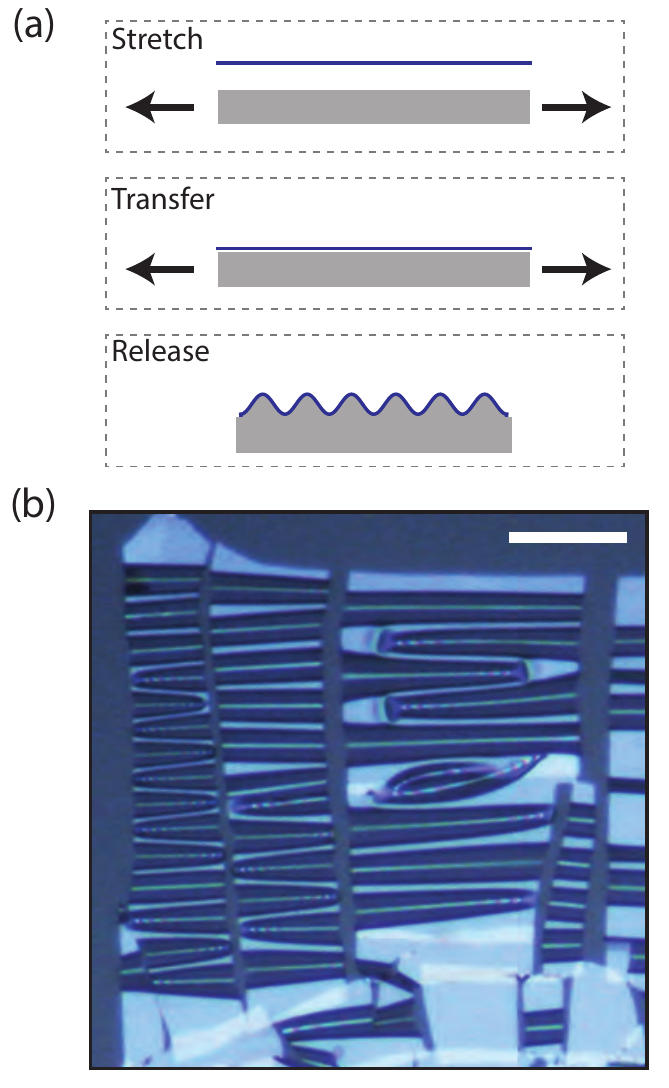}
		\caption{(a) A schematic representation of the steps used to obtained buckled 
			flake on top of PDMS. (b) An optical microscope image of buckled BSCCO flake.
			The scale bar corresponds to 12~$\mu$m.}
		\label{Fig4}
	\end{center}
\end{figure}

The Young's modulus of rigidity can also be determined by a simple buckling technique \cite{stafford_buckling-based_2004}. We use this 
technique to independently verify the AFM results. A schematic representation 
of the steps used for buckling of BSCCO is shown in Fig.~\ref{Fig4}(a). In this process, 
flake is transferred directly on a pre-stressed substrate. We use a PDMS substrate (PF-X4 6.5 mil from GelPak), which 
has Young's modulus of $E_s~=~492$~kPa and Poisson's ratio of $\nu~=~0.5$ \cite{iguiniz_revisiting_nodate}. 
PDMS is elongated up to 30-40\% of its original length in a 
direction perpendicular to its surface to generate the pre-stress.
After releasing the stress from PDMS, BSCCO flake buckles with a particular wavelength.

Fig.~\ref{Fig4}~(b) shows an optical microscope image of buckled BSCCO flake over PDMS substrate. 
Different buckling wavelengths for different thickness are evidently visible.
The wavelength of induced ripples $(\lambda_b)$ is independent of initial stress and depends 
on the elastic properties of both flake and substrate, given by \cite{stafford_buckling-based_2004}:

\begin{equation}
	\lambda_b=2\pi t \left[\frac{(1-{\nu_s}^2) E}{3(1-{\nu}^2)E_s} \right]^{\frac{1}{3}}.
	\label{eqn3}
\end{equation}

The wavelength is 
estimated by analyzing the optical microscope image of the buckled structure. For calibration of 
length, micron/pixel is calculated using a pre-patterned sample with known dimensions. 
The thickness of the flake was measured to be 7~UC using AFM.
The average value of $\lambda_b$  is $\sim$2.11~$\mu$m, calculated from four different 
data points. Using Eq.~\ref{eqn3}, we estimated the Young's modulus of rigidity to be 24.5~GPa,
which is similar to values obtained from the AFM measurement.

It is interesting to contrast our results on few UC samples to the 
observations made on the bulk crystals of BSCCO and other high-T$_c$ 
layered superconductors. The Young’s modulus of rigidity for bulk 
crystals of BSCCO has been reported over a range of values (see Table~\ref{tbl1} ). 
Crystalline quality dependent variations in the 
Young’s modulus of rigidity has been observed for other layered 
high-T$_c$ superconductors such as YBa$_2$Cu$_3$O$_7$ \cite{sihan_elastic_1989,bishop_bulk-modulus_1987}.
The reduction in modulus of rigidity and breaking strength can be 
attributed to defects formed during the crystal growth process \cite{falin_mechanical_2017,song_large_2010}. 
Under a normal applied load, the material tends to yield at the defect sites 
first, before stretching of the atomic bonds \cite{Griffith_phenomena_1921}. 
Importance of the defect density in determining elastic coefficients 
and the breaking strength has been reported for mesoscopic samples of different materials \cite{zandiatashbar_effect_2014,lopez-polin_increasing_2015,falin_mechanical_2017}. 
Layered superconductors having several weakly interacting layers with defects are therefore expected 
to show reduced material stiffness.

\begin{table*}
	\caption{Summary of Young's modulus of rigidity for high-$T_c$ superconductors}
	\label{tbl1}
	\begin{tabular}{llp{5cm}l}
		\hline
		Material & Structure & Technique & E[GPa] \\
		\hline
		BSCCO & bulk polycrystaline & ultrasonic velocity\cite{sihan_elastic_1989} & 38.8    \\
		& bulk single crystal & vibrating reed\cite{nes_anomalies_1991} & 70   \\
		& few unit cells thick & AFM and buckling methods [This work] & 22~-~30    \\
		YBa$_2$Cu$_3$O$_7$ & bulk single crystal & ultrasonic velocity\cite{bishop_bulk-modulus_1987}  & 46.4  \\
		\hline
	\end{tabular}
\end{table*}

For application towards the composite nanoelectromechanical devices, 
the resonant frequency of the mechanical resonator is an important design 
parameter. From the variation in the Young's modulus and pre-stress reported
in this study, we expect the mechanical resonance frequency to be in the 
range of 6~MHz to 18~MHz for 5~UC thick crystals of 6~$\mu$m diameter, as also
observed experimentally \cite{sahu_nanoelectromechanical_2019}. We further 
note that for few UC thick mechanical resonators, the resonant frequency is 
primarily dominated by the pre-stress induced by the exfoliation process. 
The expected high frequency of BSCCO mechanical resonators and typical linewidths 
of superconducting microwave resonators ($<$~500~kHz) place these 
devices in the sideband-resolved limit, an important criterion for experiments 
in the quantum limit \cite{poot_mechanical_2012}.

To summarize, we have studied the mechanical properties of
exfoliated thin BSCCO crystals using deformation caused by an AFM tip. 
Finite element simulations are used for the numerical analysis 
of spatial compliance maps, and to extract the Young's modulus and pre-stress.
The reported mechanical properties could potentially be useful 
in engineering nanoelectromechanical resonators of BSCCO for various applications.

\section*{SUPPLEMENTARY MATERIAL}
See supplementary material for the characterization of additional flakes and details of simulations.

\begin{acknowledgements}
	
This work was supported by STC-ISRO. VS acknowledges the fabrication facilities at CeNSE and AFM facilities at the Department of Physics, IISc-Bangalore, funded by the Department of Science and Technology. MMD acknowledges Nanomission grant SR/NM/NS-45/2016 and the Department of Atomic Energy for support.
	
\end{acknowledgements}

\end{document}